# Diffuse neutrino flux measurements with the Baikal-GVD neutrino telescope

V.A. Allakhverdyan,[1] A.D. Avrorin,[2] A.V. Avrorin,[2] V. M. Aynutdinov,[2] Z. Bardačová,[3] I.A. Belolaptikov,[1] I.V. Borina,[1] N.M. Budnev,[4] V.Y. Dik,[1,5] G.V. Domogatsky,[2] A.A. Doroshenko,[2] R. Dvornický,[3] A.N. Dyachok,[4] Zh.-A.M. Dzhilkibaev,[2] E. Eckerová,[3] T.V. Elzhov,[1] L. Fajt,[6] A.R. Gafarov,[4] K.V. Golubkov,[2] N.S. Gorshkov,[1] T.I. Gress,[4] K.G. Kebkal,[7] V.K. Kebkal,[7] A. Khatun,[3] E.V. Khramov,[1] M.M. Kolbin,[1] K.V. Konischev,[1] A.V. Korobchenko,[1] A.P. Koshechkin,[2] V.A. Kozhin,[8] M.V. Kruglov,[1] V.F. Kulepov,[9] Y.M. Malyshkin,[1] M.B. Milenin,[2] R.R. Mirgazov,[4] D.V. Naumov,[1] V. Nazari,[1] D.P. Petukhov,[2] E.N. Pliskovsky,[1] M.I. Rozanov,[10] V.D. Rushay,[1] E.V. Ryabov,[4] G.B. Safronov,[2] D. Seitova,[1,5] B.A. Shaybonov,[1] M.D. Shelepov,[2] F. Šimkovic,[3,6] A.E. Sirenko,[1] A.V. Skurikhin,[8] A.G. Solovjev,[1] M.N. Sorokovikov,[1] I. Štekl,[6] A.F. A. Stromakov,[2] O.V. Suvorova,[2] V.A. Tabolenko,[4] Y.V. Yablokova,[1] D.N. Zaborov[2]

(Baikal-GVD Collaboration)

[1] *Joint Institute for Nuclear Research, Dubna, 141980 Russia*
[2] *Institute for Nuclear Research of the Russian Academy of Sciences, Moscow, 117312 Russia*
[3] *Comenius University, Bratislava, 81499 Slovakia*
[4] *Irkutsk State University, Irkutsk, 664003 Russia*
[5] *Institute of Nuclear Physics ME RK, Almaty, 050032 Kazakhstan*
[6] *Czech Technical University, Institute of Experimental and Applied Physics, CZ-11000 Prague, Czech Republic*
[7] *EvoLogics GmbH, Berlin, 13355 Germany*
[8] *Skobeltsyn Research Institute of Nuclear Physics, Moscow State University, Moscow, 119991 Russia*
[9] *Nizhny Novgorod State Technical University, Nizhny Novgorod, 603950 Russia*
[10] *St. Petersburg State Marine Technical University, St. Petersburg, 190008 Russia*

ABSTRACT

We report on the first observation of the diffuse cosmic neutrino flux with the Baikal-GVD neutrino telescope. Using cascade-like events collected by Baikal-GVD in 2018–2021, a significant excess of events over the expected atmospheric background is observed. This excess is consistent with the high-energy diffuse cosmic neutrino flux observed by IceCube. The null cosmic flux assumption is rejected with a significance of 3.05σ. Assuming a single power law model of the astrophysical neutrino flux with identical contribution from each neutrino flavor, the following best-fit parameter values are found: the spectral index $\gamma_{astro} = 2.58^{+0.27}_{-0.33}$ and the flux normalization $\varphi_{astro} = 3.04^{+1.52}_{-1.21}$ per one flavor at 100 TeV.

## 1. INTRODUCTION

The high-energy diffuse neutrino flux observed at Earth today has been generated by neutrino emission from the entire set of neutrino sources over the time from distant cosmological epochs to the present day. Galactic and extragalactic objects are among possible sources of these high-energy neutrinos, such as, e.g., supernova remnants and active galactic nuclei [1,2]. The standard approach [3-7] used by a wide range of theoretical models that describe the generation of neutrino fluxes in cosmic ray sources suggests the production of neutrinos mainly during the decay of charged pions produced in *pp* and *pγ* interactions. In this case, the neutrino flux emitted by a source consists of neutrinos of various flavors in a proportion $\nu_e : \nu_\mu : \nu_\tau \approx 1 : 2 : 0$. Due to the effect of neutrino oscillations the flavor ratio changes with the distance to the source. Since the oscillation length is considerably smaller than the characteristic distances from the source to the detector, the flavor ratio becomes $\nu_e : \nu_\mu : \nu_\tau \approx 1 : 1 : 1$ [8, 9].

Detection of neutrinos with neutrino telescopes is achieved through detecting the Cherenkov radiation emitted by secondary particles produced in neutrino interactions. Charged current (CC) muon neutrino interactions yield long-lived muons that can pass several kilometers through the water or ice, leading to a track signature in the detector. For high-energy muon-like events the accuracy of track reconstruction is typically better than 1°. Neutral current (NC) neutrino interactions and CC interactions of electron and tau neutrinos generally yield cascades – hadronic and electromagnetic showers of charged particles. For a typical neutrino telescope, these showers are quasi point-like, however their Cherenkov radiation is highly anisotropic. Directional resolution for cascades is typically a few degrees (for sea- and lake-based experiments). An advantage of the cascade detection channel (over track detection) is its high energy resolution (10–30%) as well as a low atmospheric neutrino background. The cascade channel allows for effective measurement and characterization of the energy-dependent astrophysical neutrino flux.

IceCube discovered a diffuse flux of high-energy astrophysical neutrinos in 2013 [10]. Various IceCube datasets have been employed for the diffuse flux studies: a sample of high-energy neutrinos which includes both tracks and cascades with interaction vertices within the instrumented volume [11], a sample of up-going tracks (mostly muon neutrinos) [12], a sample of cascade-like events (mostly electron and tau neutrinos) [13], and a sample of tracks that start within the instrumented volume [14]. The flux, observed using 6 years of IceCube cascade data [13] is consistent with an isotropic single power law model with spectral index γ = 2.53 ±0.07 and a flux normalization for each neutrino flavor of $\varphi_{astro}$ = $1.66^{+0.25}_{-0.27}$ at $E_0$ =100 TeV. A mild excess of high-energy events, consistent with the IceCube's diffuse neutrino flux, has also been reported by ANTARES [15, 16] albeit with a statistical significance of under 2σ.

Baikal Gigaton Volume Detector (Baikal-GVD) is a cubic kilometer scale deep underwater Cherenkov detector currently under construction in Lake Baikal, Russia, aimed at a search for incoming neutrinos with energies between several TeV and tens of PeV [17]. The prime physics goal for Baikal-GVD is to measure and investigate the neutrino flux of astrophysical origin observed by IceCube, with different systematics and a complementary field of view. The Baikal-GVD detector is formed by sub-arrays, so-called clusters, each of which is connected to the shore station by its own electro-optical cable. Each cluster is an independent array comprising 288 light sensors – optical modules (OMs). The modular structure of the telescope allows to perform studies even at early stages of detector deployment. The first full-scale Baikal-GVD cluster was deployed in April 2016. In 2017–2022, nine additional clusters were deployed and commissioned, increasing the total number of optical modules to over 2800 OMs. The current rate of array deployment is about two clusters per year.

The Baikal Collaboration has long-term experience to search for diffuse neutrino flux via the cascade mode [18-20]. Here we report on the first measurement of the astrophysical neutrino flux using cascade-like events in Baikal-GVD data from 2018–2021.

## 2. BAIKAL-GVD NEUTRINO EXPERIMENT

The Baikal-GVD neutrino telescope is located in the Southern part of Lake Baikal (51°50′ N, 104°20′ E) at about 4 km from the shore. The lake depth at the site is 1366 m. The deployment of new clusters of the telescope is performed in the periods of strong ice cover of the lake during 7–8 weeks from the mid of February to the be- ginning of April. At wavelengths λ = 480–500 nm the light absorption length is $La$ = 21–23 m and the scattering length is $Ls$ = 60–80 m. Seasonal variations of the light absorption length normally do not exceed 5%. The light scattering in Baikal water is strongly anisotropic, with an average scattering angle cosine of about 0.9 [21, 22]. A typical count rate of background OM hits from water luminescence is about 20–40 kHz depending on the depth [23]. These background hits are typically one-photo-electron (1 p.e.) hits. The cut on the lowest allowed charge of OM hits Q > 1.5 p.e. allows for suppression of noise pulses produced by water luminescence by at least an order of magnitude.

The design and basic characteristics of the telescope data acquisition system are described elsewhere [17, 24]. Each OM comprises a 10-inch photo-multiplier tube (PMT) with high quantum efficiency of photocathode (Hamamatsu R7081-100), a high voltage unit and front-end electronics, all together enclosed in a pressure-resistant glass sphere. The OMs are attached to vertical strings, each holding 36 OMs, as well as three "Section Modules" and one "String Module". On those strings, the OMs are installed with 15 m vertical spacing between 750 and 1275 m below the surface. The Section Modules serve a group of 12 OMs providing power to the OMs and digitizing the PMT signals with a 5 ns resolution. The String Module acts as a hub for power distribution and communication with the Section Modules. Each Baikal-GVD cluster comprises 8 strings, as is shown in Fig.1. The seven outer strings are laid out at about 60 m distance from the central one. The clusters are arranged on the lake bed in a hexagonal pattern, with 250–300 m distance between the cluster centers.

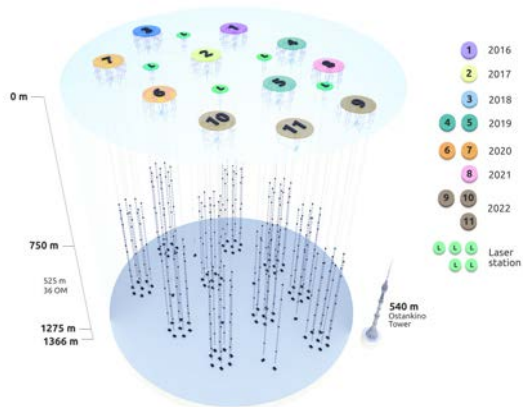

**Figure 1.** Ten Baikal-GVD clusters in the 2022 configuration. Also shown are stations with calibration laser light sources and experimental strings. The season of deployment of each cluster is shown on the right

3. DATA ANALYSIS

The search for high-energy astrophysical neutrinos using cascade-like events includes the selection and reconstruction of high-energy showers generated in neutrino interactions in the telescope detection volume. The cuts on quality variables are optimized on Monte Carlo simulations and tuned using the data sample accumulated in 2016–2017. In this analysis we used only OM hits with charge Q > 1.5 p.e. Such selection allows for substantial suppression of the noise pulses from water luminescence. For further suppression of noise pulses, we select events with a large multiplicity of triggered OMs $N_{hit} > 7$ at three or more strings, and require that hits satisfy the causality condition [24].

The procedure for reconstructing energy, direction, and vertex coordinates of high-energy showers is performed in two steps [25]. In the first step, the shower vertex coordinates $\vec{r}_{sh}$ are reconstructed by minimization of $\chi_t^2$ function using the time information from the selected hits. In this procedure, the shower is assumed to be a point-like source of light. In the second step, the shower energy ($E_{sh}$) and direction (θ, φ) are reconstructed by applying the maximum-likelihood method with the use of the shower coordinates reconstructed in the first step. Poorly reconstructed events are rejected by applying cuts on quality parameters, including the values of $\chi_t^2$ and maximum-likelihood function, OMs hit multiplicity $N_{hit}$ [19]. The precision of the reconstruction of shower energy and direction was estimated using Monte Carlo simulation of the Baikal-GVD cluster [20]. The precision of energy reconstruction substantially depends on the energy of the cascade and on its position and orientation relative to the cluster, typically varying in the range 10–30%. The precision of reconstruction of the shower direction also depends on the shower energy, position, and orientation and is 2°–4° (median value) [20].

We used Baikal-GVD Monte Carlo simulation packages to simulate the cosmic ray background with CORSIKA7.74 [26] using the proton spectrum proposed in [27]. The primary interaction of protons in the Earth atmosphere was simulated based on the SIBYLL 2.3c model [28]. The propagation of muons in water down to the detector depth was simulated based on the MUM program [29]. The efficiency of neutrino event registration was estimated by simulating the passage of neutrinos through the Earth and the interaction in the sensitive volume of the facility using the neutrino cross sections from [30, 31], the τ lepton decay cross sections from [32], and the model of the Earth profile from [33]. The telescope response to the Cherenkov radiation of showers from neutrino interactions was simulated accounting for the shower development in water, as well as light absorption, scattering and light velocity dispersion in water.

Astrophysical neutrino event selection efficiencies were tested assuming a flux with equal numbers of neutrinos and anti-neutrinos, and with an equal neutrino flavor mixture at Earth: $(\nu_e : \nu_\mu : \nu_\tau) = 1 : 1 : 1$. The one flavor (1f) flux presented by IceCube in [34] was chosen as baseline:

$$\Phi_{\nu+\bar{\nu}}^{1f} = 2.06 \times \left(\frac{E_\nu}{10^5}\right)^{-2.46} \left(\frac{1}{GeV \cdot cm^2 \cdot sr \cdot s}\right). \quad (1)$$

The conventional atmospheric neutrino flux from pion and kaon decays was modeled according to [35]. Atmospheric prompt neutrinos were simulated according to the BERSS model [36].

## 4. RESULTS AND DISCUSSION

We use Baikal-GVD data collected between April 2018 and March 2022 for the search of astrophysical neutrinos. The telescope was operating in the configuration with 3 clusters in 2018–2019, 5 clusters in 2019–2020, and 7 clusters in 2020–2021, while from April 2021 to March 2022, the telescope consisted of 8 clusters. In this study, we report on results of data analysis for individual clusters as independent setups. A sample of $3.49 \times 10^{10}$ events was collected by the basic trigger of the telescope. After applying noise hit suppression procedures, cascade reconstruction and applying cuts on reconstruction quality parameters the sample of 14328 cascades with reconstructed energy $E_{sh} > 10$ TeV and OM hit multiplicity $N_{hit} > 11$ was selected.

### *4.1 All-sky analysis*

Following the same procedure as in our previous analyses [20], high-energy cascade events with OM hit multiplicity $N_{hit} > 19$ and reconstructed energy $E_{sh} > 70$ TeV were selected and additional cuts which suppress events from atmospheric muons were applied [37]. The fraction of background events from atmospheric muons in the selected sample is expected at a level of 50%. As a result, in addition to 10 events selected from the 2018–2020 data sample [37], 6 events were selected from 2021 data. A total of 8.2±2.0 (sys.) events are expected from the simulations of the background (7.4 from atmospheric muons and 0.8 from atmospheric neutrinos) and 5.8 events are expected from the astrophysical best-fit flux derived in this work (see subsection *4.2*). The effect of the uncertainty of the detector response on signal and background is evaluated by varying input parameters in the Monte Carlo simulations. The light absorption length uncertainty is about ±5% and the optical module sensitivity varies within ±10%. Also a ±15% uncertainty on the normalization of the conventional atmospheric neutrino component is considered [35]. The uncertainties coming from independent sources are added in quadrature in the overall estimation. Taking into account the systematic effects according to the method of [38], the significance of the excess was estimated to be 2.22σ with the null-cosmic hypothesis rejected at 97.36% confidence level. The distributions of the reconstructed cascade energies and zenith angles are shown in Fig.2 (black points). Also shown are distributions for Monte Carlo simulations with the signal and background contributions. The Monte Carlo simulated histograms are stacked (filled colors). The energy and zenith distributions of data are consistent with expectations for the baseline (IceCube) flux of neutrinos of astrophysical origin (1). It should be noted that this set of data included an event with the energy on the order of 1 PeV. This was the first event with energy of such scale, which was selected from the Baikal-GVD data. The null-cosmic hypothesis is rejected at 95.9% confidence level (2.0σ significance of excess) for such event detection. Three events in the all-sky analysis were reconstructed as cascades from below the horizon. Reconstructed parameters of the 16 selected events are shown in Table 1.

### *4.2 Upward-going cascade analysis*

Restricting the analysis to upward-going directions allows for effective suppression of the atmospheric muon background, thus improving the neutrino sample purity and enabling the extension of the analysis towards lower energies. Cascade-like events with reconstructed energy $E_{sh} > 15$ TeV, OM hit multiplicity $N_{hit} > 11$ and reconstructed zenith angle $\cos\theta < -0.25$ were selected as astrophysical neutrino candidates. Total of 11 events have been selected from 2018−2021 data sample, while 3.2±1 atmospheric background events are expected (2.7 from atmospheric conventional and prompt neutrinos and 0.5 events from mis-reconstructed atmospheric muons). Taking into account the systematic effects (see subsection *4.1*) according to the method of [38], the significance of the excess was estimated to be 3.05σ with the null-cosmic hypothesis rejected at 99.76% C.L. The parameters of the 11 upward-going cascades are shown in Table 2. The median value of the error in the reconstruction of the cascade direction varies from 1.9° to 5.4°. The two events with the highest reconstructed energies, 91 TeV and 224 TeV, are also present in the all-sky high-energy sample discussed earlier.

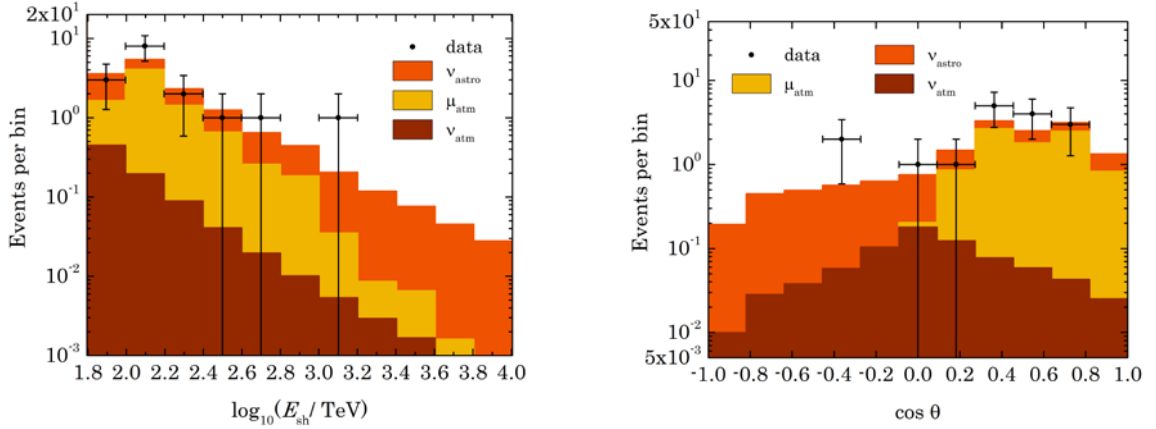

**Figure 2.** Reconstructed cascade energy (left panel) and zenith (right panel) distributions obtained in the all-sky analysis. Black points are data, with statistical uncertainties. The colored bands show the expected contribution from background atmospheric neutrinos (brown) and atmospheric muons (yellow), as well as from the best fit astrophysical neutrino flux obtained in this work (orange).

The measured 11 events and the expected number of background events have been analyzed to characterize the diffuse astrophysical neutrino flux. We parameterize the isotropic diffuse astrophysical neutrino flux $\Phi_{astro}^{\nu+\bar{\nu}}$ in the single power law model assuming equal numbers of neutrinos and anti-neutrinos and equal neutrino flavors at Earth. The model is characterized by spectral index $\gamma_{astro}$ and normalization $\varphi_{astro}$ of the one-flavor neutrino flux in units of GeV$^{-1}$cm$^{-2}$sr$^{-1}$s$^{-1}$:

$$\Phi_{astro}^{\nu+\bar{\nu}} = 3 \times 10^{-18} \varphi_{astro} \left(\frac{E_\nu}{E_0}\right)^{-\gamma_{astro}}, \qquad (2)$$

where $E_0 = 100$ TeV. The best fit parameters for the observed data are determined by a binned likelihood approach. In this procedure, the data sample is binned in reconstructed shower energy. The observed count $n_i$ in each bin $i$ is compared to a model that predicts the mean count rate $\lambda_i$ in each bin using a Poisson likelihood function:

$$L = \prod_{i=1}^{N} \frac{e^{-\lambda_i} \lambda_i^{n_i}}{n_i!} . \qquad (3)$$

The expected rates $\lambda_i$ are composed by astrophysical neutrinos and background events of atmospheric muons and atmospheric neutrinos. Based on the Poisson likelihood function (3) the following test statistic (TS) is used to compare the observed event counts with the Monte Carlo model predictions:

$$TS = -2\ln L + \sum_{k=1}^{} \left(\frac{(g_k - g_k^0)}{\sigma(g_k)}\right)^2 . \qquad (4)$$

The second term in (4) accounts for the systematic uncertainties discussed above which are incorporated in the test statistic as nuisance parameters in form of Gaussian distributions of prior $g_k$ and width deviation $\sigma(g_k)$ from central value $g_k^0$. A maximum-likelihood method is applied to find the best-fit values of $\gamma_{astro}$ and $\varphi_{astro}$ by varying these parameters until (4) is minimized. We find the best-fit parameters as following: the spectral index $\gamma_{astro} = 2.58^{+0.27}_{-0.33}$ and the flux normalization for each neutrino flavor at $E_0 = 100$ TeV $\varphi_{astro} = 3.04^{+1.52}_{-1.27}$.

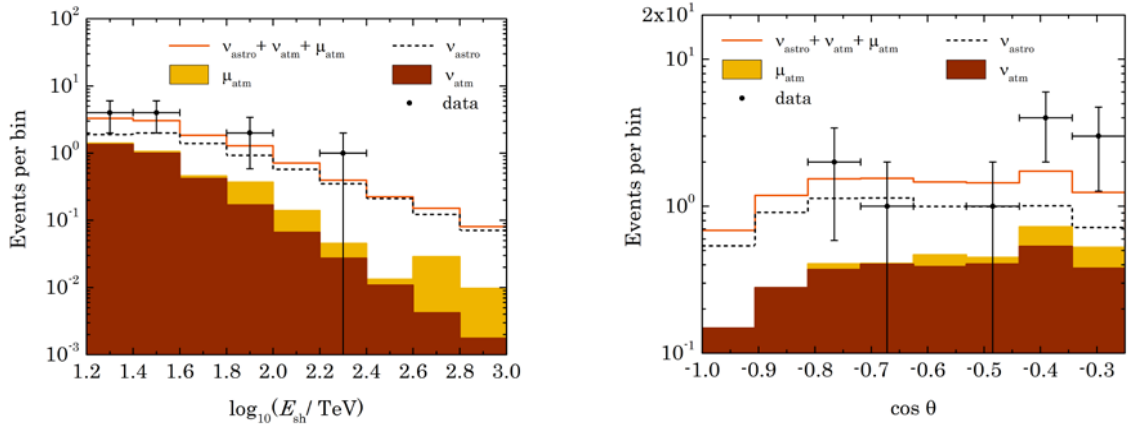

**Figure 3.** Reconstructed cascade energy (left panel) and zenith (right panel) distributions obtained in the upward-going cascade analysis. Black points are data, with statistical uncertainties. The best-fit distribution of astrophysical neutrinos (dashed line), expected distributions from atmospheric muons (yellow) and atmospheric neutrinos (brown) and the sum of the expected signal and background distributions (orange line) are also shown. The atmospheric background histograms are stacked (filled colors).

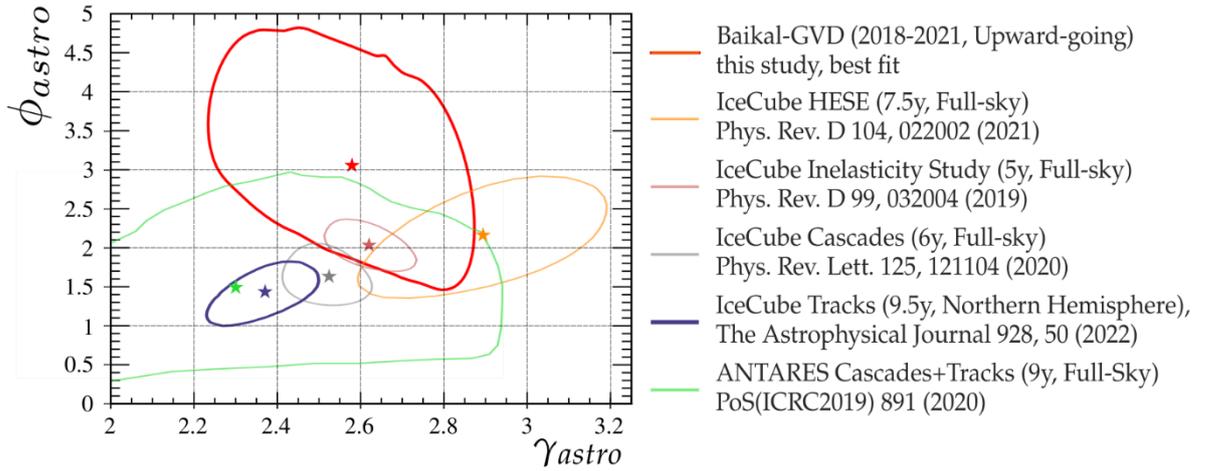

**Figure 4.** The best fit parameters and the contours of the 68% confidence region (red curve) for the single power law hypothesis obtained in the upward-going cascade analysis of the Baikal-GVD data. Other best fits are shown for studies based on high-energy starting events (orange curve) [11], cascade-like events (gray curve) [13], an inelasticity study (purple curve) [14] and track-like events (blue curve) [12] by IceCube and ANTARES observation in a combined study of tracks and cascades (green curve) [16].

The energy and zenith distributions of the 11 events are shown in Fig.3 together with the distributions obtained by Monte Carlo simulation. The atmospheric background histograms are stacked (filled colors). The best-fit parameters and 68% C.L. contours for this cascade analysis together with the results from other neutrino telescopes [11-16] are shown in Fig 4. The Baikal-GVD upward-going neutrino (cascades) measurements are consistent with the IceCube measurements (except muon neutrino sample [12]) and the ANTARES all-neutrino flavor measurements.

*4.3 Baikal-GVD sky map*

Figure 5 shows the reconstructed sky-map positions and the uncertainty regions of the cascade events selected in the all-sky analysis (solid circles) and the upward-going cascade analysis (dashed circles). The two upward-going events which are common to both the data samples (GVD190523CA and GVD210418CA) are shown as dashed circles. Note that about half of the events are background from atmospheric muons and atmospheric neutrinos. The circles around events correspond to detection probabilities of 90% for each event. In general, the cascade events appear to be distributed somewhat isotropically all over the sky, consistent with the dominance of extragalactic sources. Note that source population studies with cascade events can be challenging due to large angular uncertainty. However, dedicated searches of correlation of arrival directions of the most energetic events with known neutrino sources are possible. The strongest neutrino candidate source in extragalactic sky at $E > 200$ TeV according to IceCube is TXS 0506+056 [39]. It is curious to note that the arrival direction of the highest energy upward-going neutrino candidate event in the Baikal-GVD data (GVD210418CA, $E = 224$ TeV) is consistent with TXS 0506+056. A combined analysis with radio data shows that the neutrino events from TXS 0506+056 are correlated with the radio activity of this source: a full analysis of the event is presented in dedicated publication [40].

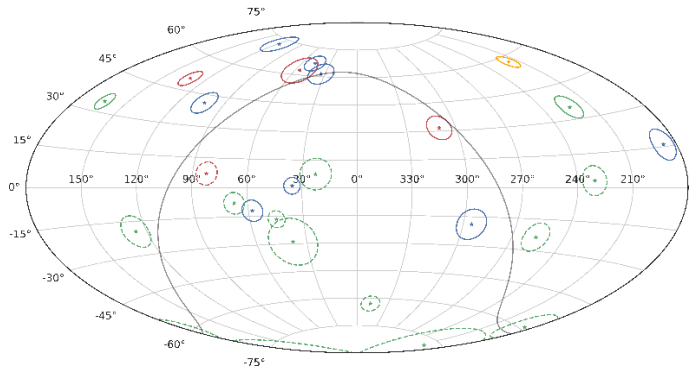

**Figure 5.** The Baikal-GVD high-energy cascade sky map (in equatorial coordinates). The best-fit positions and 90% angular uncertainty regions are shown. Dashed circles show under-horizon events (selected in the upward-going analysis, subsection *4.2*), while solid circles represent events above horizon (selected in the all-sky analysis, excluding upward-going events). Colour represents energy of the events: green is below 100 TeV, blue is between 100 TeV and 200 TeV, red is between 200 TeV and 1000 TeV, and orange is above 1 PeV. The Galactic plane is indicated as a grey curve.

As can be seen in Fig.5, the arrival directions of 3 out of 25 events are consistent with the Galactic plane. Moreover, the 90%-errors of two of them (events GVD190216CA and GVD190604CA in Table 1) intersect at a region containing the well-known gamma-ray active binary system LS I +61°303 [41, 42]. It is known to be variable, with flares coming periodically in a wide range of the electromagnetic spectrum from radio to TeV. Notice that, while a continued neutrino emission from this source is excluded by IceCube observations, a flaring source cannot be definitely ruled out. The chance probability to observe such a doublet near LS I +61°303 due to background events was estimated as 0.007 (2.7σ significance), not including the look-elsewhere effect.

## CONCLUSION

We presented the first measurements of the astrophysical neutrino flux using samples of cascade events collected by the Northern Hemisphere neutrino telescope Baikal-GVD in 2018–2021. Two analyses were performed using cascade events. In the first case the sample of high-energy cascades with $E_{sh} > 70$ TeV from all sky directions was analyzed. A total of 16 events were selected as astrophysical neutrino candidates while 8.2 events were expected from atmospheric background. The significance of the excess was estimated to be 2.22σ with the null-cosmic hypothesis rejected at 97.36% confidence level.

The second analysis used a sub-sample of upward moving cascades with energy $E_{sh} > 15$ TeV. A total of 11 events have been selected as astrophysical neutrino candidates, while 3.2 ±1 atmospheric background events are expected. The significance of the excess over the expected number of atmospheric background events was estimated as 3.05σ. We have made a global fit to these neutrino data, fitting the cascade energy distribution, to extract information about the astrophysical neutrino flux. The measured values of an astrophysical power law spectral index of $\gamma_{astro} = 2.58^{+0.27}_{-0.33}$ and the flux normalization for each neutrino flavor at $E_0 = 100$ TeV $\varphi_{astro} = 3.04^{+1.52}_{-1.27}$ are in good agreement with the previous fits derived in various analyses of the IceCube data and ANTARES data. With these results we, for the first time, confirm the IceCube observation of astrophysical diffuse neutrino flux with 3σ significance.

**Table 1.** Parameters of 16 high-energy cascade events selected in the all-sky analysis: date of observation as Modified Julian Date, reconstructed energy, zenith angle, Galactic longitude and latitude, right ascension and declination, 50% and 90%-containment angular uncertainty region, distance between shower vertex and central string of cluster. The event name (left column) encodes the event detection date in the format yymmdd.

| Event name | MJD | $E_{sh}$ | $\vartheta$ | $l$ | $b$ | RA | Dec | 50% unc. | 90% unc. | $\rho$ |
|---|---|---|---|---|---|---|---|---|---|---|
|  |  | TeV | deg. | deg. | deg. | deg. | deg. | deg. | deg. | meter |
| GVD181010CA | 58401.77863426 | 105 | 37 | 142.6 | 30.4 | 118.2 | 72.5 | 2.3 | 4.5 | 70 |
| GVD181024CA | 58415.88952546 | 115 | 73 | 164.1 | −54.4 | 35.4 | 1.1 | 2.5 | 4.5 | 90 |
| GVD190216CA | 58530.03428241 | 398 | 64 | 141.4 | 5.8 | 55.6 | 62.4 | 3.3 | 6.9 | 101 |
| GVD190517CA | 58620.31961806 | 1200 | 61 | 99.9 | 54.9 | 217.7 | 57.6 | 2.0 | 3.0 | 96 |
| GVD190523CA | 58626.44462963 | 91 | 109 | 200.4 | −58.4 | 45.1 | −16.7 | 2.2 | 4.5 | 49 |
| GVD200117CA | 58865.65752315 | 83 | 50 | 190.0 | 64.0 | 163.6 | 34.2 | 2.1 | 3.3 | 73 |
| GVD190604CA | 58638.82969907 | 129 | 50 | 132.7 | 0.1 | 33.7 | 61.4 | 3.5 | 5.5 | 52 |
| GVD200826CA | 59087.58636574 | 110 | 71 | 21.0 | −19.2 | 295.3 | −18.9 | 2.0 | 7.9 | 84 |
| GVD201222CA | 59205.54451389 | 74 | 92 | 58.3 | 63.1 | 223.0 | 35.4 | 1.8 | 5.1 | 19 |
| GVD210117CA | 59231.02799769 | 246 | 57 | 168.8 | 38.8 | 131.9 | 50.2 | 1.6 | 3.6 | 80 |
| GVD210409CA | 59313.79668981 | 263 | 60 | 73.3 | −6.1 | 310.0 | 31.7 | 3.3 | 6.3 | 76 |
| GVD210418CA | 59322.94855324 | 224 | 115.5 | 196.8 | −14.6 | 82.4 | 7.1 | 3.0 | 5.8 | 70 |
| GVD210515CA | 59349.73187500 | 120 | 80.2 | 175.2 | 17.9 | 103.4 | 41.2 | 2.8 | 5.2 | 68 |
| GVD210716CA | 59411.42329861 | 110 | 58.7 | 135.5 | 7.1 | 46.0 | 66.7 | 2.1 | 4.1 | 93 |
| GVD210906CA | 59464.98151620 | 138 | 67.7 | 202.2 | −45.3 | 57.8 | −12.0 | 2.0 | 5.6 | 98 |
| GVD220221CA | 59631.60434028 | 120 | 67.7 | 276.9 | 77.5 | 187.2 | 15.8 | 3.2 | 5.8 | 62 |

**Table 2.** Parameters of 11 under horizon cascade events: date of observation as Modified Julian Date, reconstructed energy, zenith angle, Galactic longitude and latitude, right ascension and declination, 50% and 90%-containment angular uncertainty region, distance between shower vertex and central string of cluster. The event name (left column) encodes the event detection date in the format yymmdd.

| Event name | MJD | $E_{sh}$ | $\vartheta$ | $l$ | $b$ | RA | Dec | 50% unc. | 90% unc. | $\rho$ |
|---|---|---|---|---|---|---|---|---|---|---|
|  |  | TeV | deg. | deg. | deg. | deg. | deg. | deg. | deg. | meter |
| GVD180504CA | 58242.5739004 | 25.1 | 111.7 | 299.1 | 3.6 | 185.4 | −59.0 | 3.9 | 6.9 | 28 |
| GVD190523CA | 58626.44462963 | 91.0 | 109.0 | 200.4 | −58.4 | 45.1 | −16.7 | 2.2 | 4.5 | 49 |
| GVD200614CA | 59014.27202546 | 39.8 | 144.1 | 359.3 | 10.6 | 256.2 | −23.6 | 3.4 | 6.8 | 108 |
| GVD201112CA | 59165.01353009 | 24.5 | 136.1 | 305.0 | −15.1 | 202.2 | −77.8 | 5.4 | 11.8 | 66 |
| GVD210418CA | 59322.94855324 | 224 | 115.5 | 196.8 | −14.6 | 82.4 | 7.1 | 3.0 | 5.8 | 70 |
| GVD210501CA | 59335.45576389 | 64.6 | 112.3 | 223.4 | −67.7 | 38.1 | −28.9 | 2.6 | 12.6 | 109 |
| GVD210506CA | 59340.34252315 | 21.9 | 114.2 | 5.9 | 46.7 | 230.6 | 3.1 | 2.8 | 6.6 | 30 |
| GVD210710CA | 59405.56907407 | 24.5 | 115.5 | 139.8 | −54.2 | 22.7 | 7.4 | 3.6 | 8.6 | 83 |
| GVD210803CA | 59429.58071759 | 20.9 | 136.9 | 321.0 | −50.3 | 347.0 | −63.0 | 1.9 | 4.1 | 41 |
| GVD220121CA | 59600.45934028 | 30.9 | 110.5 | 241.3 | 10.4 | 126.2 | −19.5 | 3.4 | 7.1 | 49 |
| GVD220308CA | 59646.14655093 | 36.3 | 105.0 | 203.2 | −35.2 | 67.3 | −8.0 | 2.5 | 5.6 | 37 |

ACKNOWLEDGMENTS

We acknowledge Dmitri Semikoz, Sergey Troitsky, Yury Kovalev, Aleksandr Plavin and Nikita Kosogorov for fruitful discussion and partnership. An important role in data processing was played by the possibility of using the JINR cloud computing infrastructure. This work is supported in the framework of the State project "Science" by the Ministry of Science and Higher Education of the Russian Federation under the contract 075-15-2020-778. The work is partially supported by the European Regional

Development Fund-Project "Engineering applications of microworld physics" (CZ 02.1.01/0.0/0.0/16 019/0000766) and by the VEGA Grant Agency of the Slovak Republic under Contract No. 1/0607/20.